\documentclass[sigconf,authorversion,nonacm]{acmart}
\usepackage{cleveref}
\usepackage[inline]{enumitem}
\usepackage{tablefootnote}
\usepackage{pifont}
\usepackage{wasysym}
\usepackage{multirow}

\newcommand{\circled}[1]{\textcircled{\raisebox{-0.5pt}{$\:\!\!$\sf#1}}}

\begin{document}
\title{Mining Bug Repositories for Multi-Fault Programs}
\author{Dylan Callaghan}
\affiliation{%
  \institution{Stellenbosch University}
  \city{Stellenbosch}
  \country{South Africa}
}
\email{21831599@sun.ac.za}
\author{Bernd Fischer}
\affiliation{%
  \institution{Stellenbosch University}
  \city{Stellenbosch}
  \country{South Africa}
}
\email{bfischer@sun.ac.za}

\begin{abstract}
Datasets such as Defects4J and BugsInPy that contain bugs from real-world software projects
are necessary for a realistic evaluation of automated debugging tools.
However these datasets largely identify only a single bug
in each entry, while real-world software projects
(including those used in Defects4J and BugsInPy) typically contain multiple
bugs at the same time. We lift this limitation and describe an extension to these
datasets in which multiple bugs are identified in individual entries.
We use test case transplantation and fault location translation,
in order to expose and locate the bugs, respectively. We thus provide datasets
of true multi-fault versions within real-world software projects,
which maintain the properties and usability of the original datasets.
\end{abstract}
\maketitle
\section{Introduction}

Fault localization and program repair tools are typically evaluated over bug
repositories such as Defects4J \cite{defects4j} or BugsInPy \cite{bugsinpy}.
These repositories contain faulty program versions and their corresponding
fixes and regression test suites, which have been mined from the full
version history of multiple open-source Java and Python projects, respectively.
However, both Defects4J and BugsInPy overwhelmingly only identify a
\emph{single fault} in each faulty program version: the textual difference
between faulty and fixed versions is small and focused (typically only on a
single line), and the fixed versions pass all tests in the regression test
suites.

This single-fault nature limits the usefulness of these bug repositories
as evaluation and training data sets.  Real-world projects (including, in fact,
even those used in Defects4J and BugsInPy) often contain multiple faults
that can interact with and mask each other and thus make fault localization and
repair harder; the use of single-fault evaluation datasets thus introduces a
substantial threat to the validity of the evaluation itself. Similarly, using
these bug repositories as training data can introduce bias into learning-based
tools such as GRACE \cite{grace}.

In this paper, we describe the construction of true \emph{multi-fault} variants
of Defects4J and BugsInPy. More specifically, we describe how we identify
additional, \emph{already existing} faults in the program versions, through a
mining process based on \emph{test case transplantation} and \emph{fault
location translation}.

Test case transplantation copies tests from the regression test suite
of a given bug repository entry to an earlier entry, and checks whether
they fail there; if so, this is taken as evidence that the fault fixed in the
later program version is already present in its earlier version.
Test case transplantation was introduced by An et al.~\cite{multid4j} for the
Java-based Defects4J bug repository. We demonstrate here that it can also be
applied to the Python-based BugsInPy; however, the ``Pythonic''
programming style used in the underlying projects (e.g., the lack of explicit
export interfaces and the corresponding structure of the import clauses)
requires a substantially more complex test case extraction step to allow a
successful transplantation.

Test case transplantation only indicates that multiple faults may be present
but gives no indication where exactly they are located in the different program
versions. Since this information is required for the evaluation of tasks such
as fault localization, we complement the test case transplantation step by a
fault location translation step. This traces the identified fault locations
through the versions in the underlying \emph{project} respository back to the
version in the bug repository identfied through the test case transplantation.

We applied our technique to Defects4J v1.0.1, and to
the current version of BugsInPy.
%
%
On average, we identified
9.2 faults in each of the 311 versions of the 5 projects in Defects4J
also used by An et.\ al.~\cite{multid4j}, and
18.6 faults in 501 versions of the 17 projects in BugsInPy.
%
The identification of these faults requires one to two test cases on average to be
transplanted per fault.

\section{Background}\label{sec:background}
\begin{figure*}[t]
  \centering
  \includegraphics[width=\textwidth]{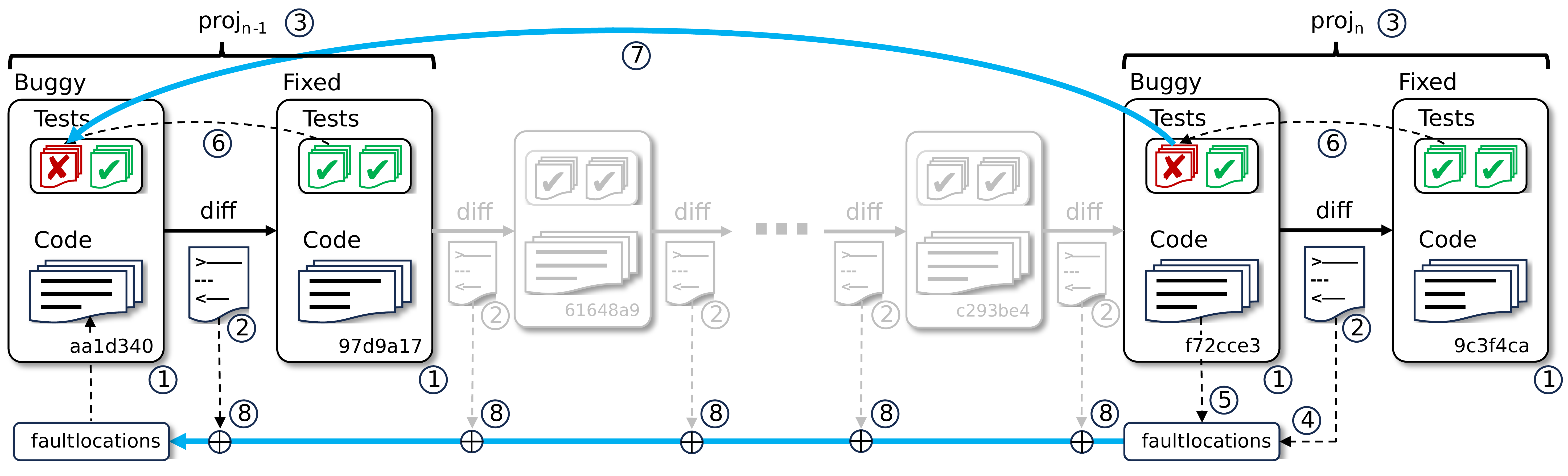}
  \caption{
  Project layout in original Defects4J~\cite{defects4j} and
  BugsInPy~\cite{bugsinpy} datasets, and construction of multi-fault variants.
  }
  \label{fig:big_picture}
\end{figure*}

\subsection{Original datasets}\label{sec:orig}

Our datasets are based on the original Defects4J \cite{defects4j} and
BugsInPy \cite{bugsinpy} datasets, which contain collections of versions
extracted or reconstructed from the original repositories of different open-source Java and
Python projects, respectively.  \Cref{fig:big_picture} shows the common
structure of all of these datasets.

Each underlying project version $v_i=(p_i, T_i)$ consists of the source code
$p_i$ and test suite $T_i$ \circled{1}.  Between any two consecutive versions
$v_{i-1}$ and $v_{i}$ in the project history, there exists a set of
changes or \emph{diff} $\Delta_i$ \circled{2} for both
the source code
and the test suite
such that applying the diff to the older version $v_{i-1}$ will produce exactly
the newer version $v_i$, i.e., $\Delta_i(v_{i-1})=v_{i}$.
%

Each bug repository entry $e$ \circled{3} references two
consecutive project versions $(v_b, v_f)$.  The
``buggy'' version $v_b$ contains a single fault exposed by at least
one failing test $t\in T_f$ from the ``fixed'' version $v_f$; this fault is repaired in
$v_f$ and all tests in $T_f=\Delta_f(T_b)$ pass.

The original datasets guarantee three properties that are important for their
use as fault localization and program repair benchmarks.
First, each fault
is \emph{exposed} by a failing test in the
buggy version's test suite. Second, each fault is \emph{repaired} in the fixed
version, and all tests in the corresponding test suite pass. Third, each diff
is \emph{minimal}, i.e., any smaller change is not a repair.

Exposure through failing tests is
the only indication of program
failure; it is necessary for spectrum-based fault localization tools,
which cannot predict faulty source code locations without failing tests.  The
fixed versions' test suites serve as specifications for program repair tools,
and the locations affected by the minimal repairs are taken as fault locations \circled{4},
and used to determine the performance of any debugging tool in either locating
or fixing the faults.
However, the diff only approximates the fault location; this
may be improved by manually constructing the fault location oracle from inspection
of the source code and bug fixing diff \circled{5} \cite{d4j_bug_identify}.

\subsection{Original dataset construction}

BugsInPy identifies the project versions $(v_b, v_f)$ referenced in
an entry $e$ by first inspecting the commit message related to the diff
$\Delta_f$ for bugfix-related terms such as ``fix''. It then checks for tests
$t_j\in T_f$ that pass in the fixed version $v_f$ but fail if they are
added to the buggy version $v_b$ (by applying the diff
\raisebox{0pt}[0pt][0pt]{$\Delta_f$} to the $v_b$'s test suite), to
ensure exposure of the bug. The addition of these test cases \circled{6} to the
buggy version changes its test suite $T_b$, but the
tests are already part of the project history,
and the code $p_b$ is identical to the repository version.
We align the version numbering in our multi-fault variant with the
commit dates, and re-label if necessary.

Defects4J also
inspects the commit messages of $\Delta_f$ for bugfix-related terms to
identify the versions $(v_b, v_f)$. However, while BugsInPy
only considers bug fixes that are already minimal, Defects4J also
selects bug fixes that contain feature additions.  It separates the
minimal bug fix
\raisebox{0pt}[0pt][0pt]{${\Delta'}_{\!\!f}$} from $\Delta_f$ to ensure minimality,
and applies the inverse
\raisebox{0pt}[0pt][0pt]{${{\Delta'}_{\!\!f}}^{-1}$} to the fixed version $v_f$
to reconstruct the ``clean'' buggy version $v_b$.
The test suite $T_f$ is then added to the buggy version $v_b$ using
\raisebox{0pt}[0pt][0pt]{${{\Delta'}_{\!\!f}}$}, similar to BugsInPy.
Hence, the buggy version $v_b$ contained in Defects4J can differ from
the referenced project version contained in the project history, however these
differences are only in the feature additions contained within $\Delta_f$.
\subsection{Related datasets}
Most fault identification datasets such as Defects4J and BugsInPy contain
program versions with only a single fault each.
The Software Infrastructure Repository (SIR)~\cite{SIR} contains a variety of
faulty programs written in multiple programming languages; of these,
\texttt{space} \cite{space}, an interpreter for an array definition language
which contains 33 real-world
single fault versions, and the Siemens set of small programs written in
C which have been seeded with single faults, are widely used for evaluation.
More recent work includes the HasBugs~\cite{hasbugs} dataset of 25 single-fault
Haskell program versions.
%
Note that these datasets are are sometimes (incorrectly) considered to be
multi-fault datasets, due to the existence of
multi-hunk faults.  Our datasets are, in contrast, proper multi-fault datasets.

True multi-fault datasets are limited, and usually either contain
synthetic or transplanted faults.
Högerle et. al.~\cite{parallel:steimann} construct a dataset of 75000 Test
Coverage Matrices (TCMs) from 15 open source Java projects. Each project version
initially contained a passing test suite, and between 1 and 32
synthetic faults were automatically injected, causing at least one test case to
fail.
%
An et al.\ partially construct a multi-fault dataset with 311 versions from the
Defects4J dataset, where the faults are exposed through the transplantation of a
failing test case, but are not all identified (i.e. indication of source code in
the version responsible for the fault). We build upon the work of An et al.\ in
this paper to construct a full multi-fault dataset from Defects4J.
Zheng et al.\ also constructed multi-fault datasets with 46 versions from the
Defects4J dataset and 217 versions from the programs contained in the
SIR~\cite{DBLP:journals/jss/ZhengWFCY18}, by manually transplanting faults from
\emph{older} versions to \emph{newer} versions in the dataset. Their technique therefore
alters the source code of underlying versions in the project history.

\section{Dataset description and statistics}

\begin{table*}
  \centering
  \begin{tabular}{|l|r|rrr|rrr|r|r|r|}
    \hline
    & & \multicolumn{3}{|c|}{Program size (loc)} & \multicolumn{3}{|c|}{Existing
    tests} & Added & \multicolumn{1}{|c|}{Drop} &
    \multicolumn{1}{|c|}{\multirow{2}{*}{$\diameter_{BPV}$}}\\
    Project      & N        & Min    & Mean      & Max     & Min  & Mean   & Max
    & \multicolumn{1}{|c|}{tests} & rate (\%) & \\
    \hline
    Chart        \cite{chart}        & 20       & 203303 & 208700.9  & 232364  &  1584 &  1752.9 &  2183 & 11.9 & 0.0  & 4.3  \\
    Lang         \cite{lang}         & 61       & 48029  & 53116.4   & 61093   &  1605 &  1872.7 &  2670 & 10.1 & 0.0  & 7.6  \\
    Math         \cite{math}         & 104      & 30521  & 121701.8  & 185273  &  880  &  2856.6 &  5187 & 7.7  & 0.0  & 5.7  \\
    Time         \cite{time}         & 23       & 70198  & 77992.2   & 99183   &  3787 &  3913.8 &  4002 & 24.0 & 0.0  & 9.2  \\
    Closure      \cite{closure}      & 103      & 99385  & 208393.9  & 269152  &  1629 &  6699.1 &  7588 & 28.9 & 1.4  & 14.9 \\
    \hline
    \textbf{Total}                   & 311      & 30521  & 133981.0  & 269152  &  880  &  3419.0 & 7588  & 16.5 & 0.3  & 8.3  \\
    \hline
    \hline
    PySnooper    \cite{pysnooper}    & 3        & 335    & 560.3     & 673     &     5 &    17.0 &    29 & 0.0  & 0.0  & 1.0  \\
    ansible      \cite{ansible}      & 18       & 101706 & 1124664.3 & 1590076 &  3101 &  7984.1 & 11020 & 5.0  & 0.0  & 5.5  \\
    black        \cite{black}        & 23       & 5241   & 66510.7   & 96049   &    18 &    81.0 &   129 & 5.1  & 0.7  & 5.9  \\
    cookiecutter \cite{cookiecutter} & 4        & 1258   & 1828.8    & 2049    &   156 &   251.5 &   298 & 1.8  & 0.3  & 2.0  \\
    fastapi      \cite{fastapi}      & 16       & 2839   & 4172.4    & 4954    &   179 &   572.0 &   793 & 3.8  & 9.8  & 3.2  \\
    httpie       \cite{httpie}       & 5        & 775    & 3106.2    & 3911    &    17 &   146.4 &   232 & 1.8  & 27.3 & 2.2  \\
    keras        \cite{keras}        & 45       & 36600  & 39474.9   & 42438   &   158 & 24817.2 & 45484 & 5.6  & 4.3  & 5.2  \\
    luigi        \cite{luigi}        & 33       & 14185  & 20071.3   & 28751   &   549 &   973.9 &  1581 & 5.0  & 26.9 & 4.4  \\
    matplotlib   \cite{matplotlib}   & 30       & 118312 & 120706.2  & 123290  &  7542 &  7814.3 &  8191 & 5.6  & 8.2  & 6.1  \\
    pandas       \cite{pandas}       & 169      & 159369 & 161675.2  & 164785  & 50989 & 63559.8 & 88768 & 59.0 & 6.0  & 45.3 \\
    sanic        \cite{sanic}        & 5        & 5506   & 7121.2    & 7604    &   638 &   641.3 &   644 & 1.0  & 0.0  & 2.0  \\
    scrapy       \cite{scrapy}       & 40       & 15636  & 20352.8   & 22631   &   923 &  1377.0 &  2050 & 16.0 & 52.6 & 8.3  \\
    spacy        \cite{spacy}        & 10       & 94575  & 97907.0   & 104284  &  1647 &  2398.4 &  2617 & 1.1  & 0.0  & 2.1  \\
    thefuck      \cite{thefuck}      & 32       & 1636   & 3679.5    & 6248    &   283 &  1087.5 &  1716 & 7.1  & 64.6 & 4.1  \\
    tornado      \cite{tornado}      & 16       & 21167  & 22957.9   & 24422   &    16 &    19.6 &    23 & 1.8  & 5.0  & 2.5  \\
    tqdm         \cite{tqdm}         & 9        & 655    & 2348.0    & 3229    &    14 &    61   &    91 & 0.9  & 21.4 & 1.6  \\
    youtube-dl   \cite{youtube-dl}   & 43       & 20515  & 82597.7   & 137957  &   324 &  1530.2 &  2365 & 6.3  & 16.2 & 6.0  \\
    \hline
    \textbf{Total}                   & 501      & 335    & 104690.3  & 1590076 &    5  &  6666.6 & 88768 & 7.5  & 14.3 & 6.3  \\
    \hline
  \end{tabular}
  \caption{Dataset statistics. $N$ is the number of versions in the project,
    program and test suite sizes are averaged over all project versions,
    $\diameter_{BPV}$ is the average number of bugs available in the multi-fault
    versions of the project.}
  \label{tab:stats}
\end{table*}

\begin{figure*}[ht]
  \centering
  \includegraphics[width=\textwidth]{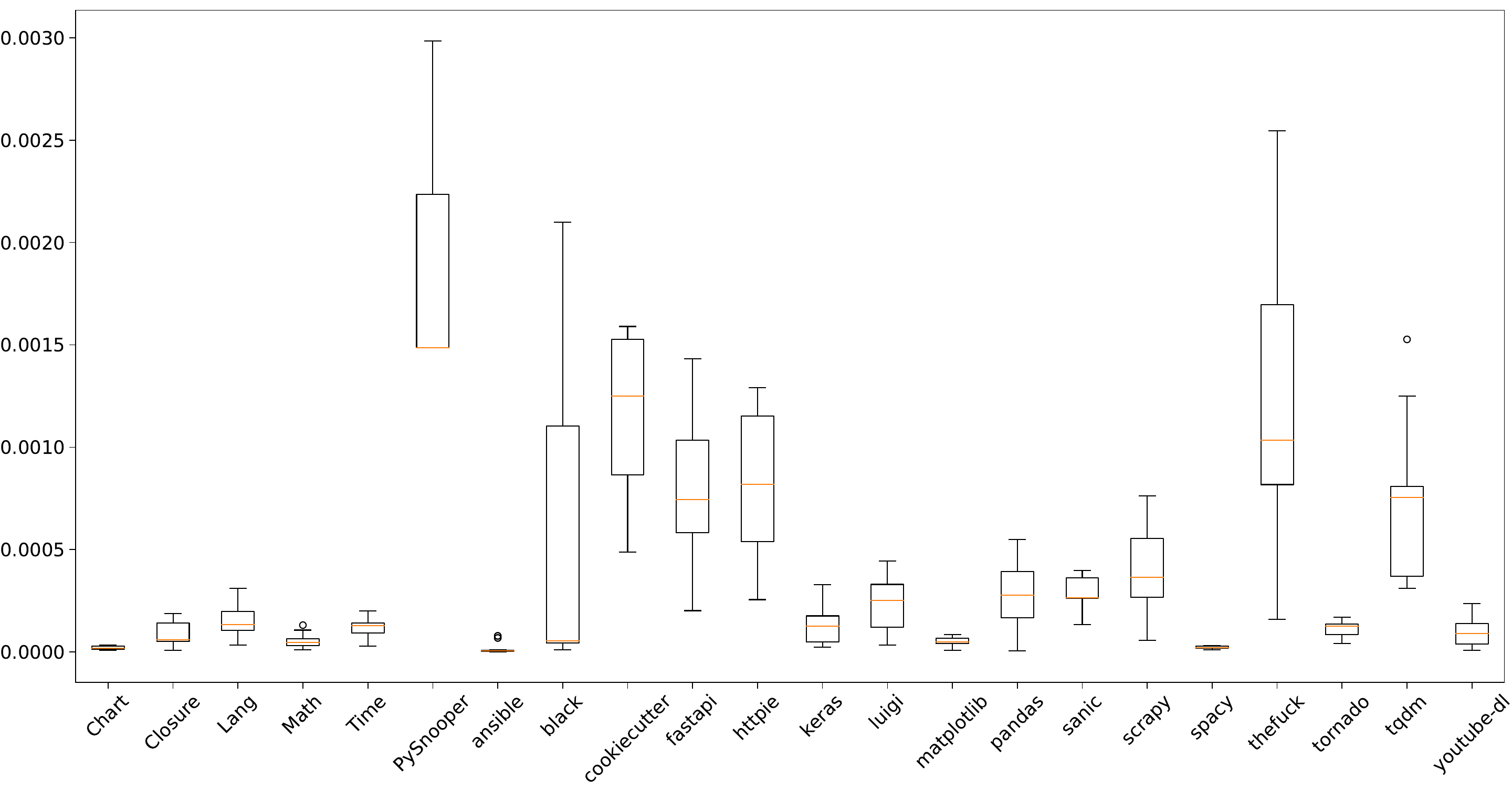}
  \caption{Average number of bugs per version, normalized by the program size of the version.}
  \label{fig:bug_stats}
\end{figure*}

\begin{figure*}[ht]
  \centering
  \includegraphics[width=\textwidth]{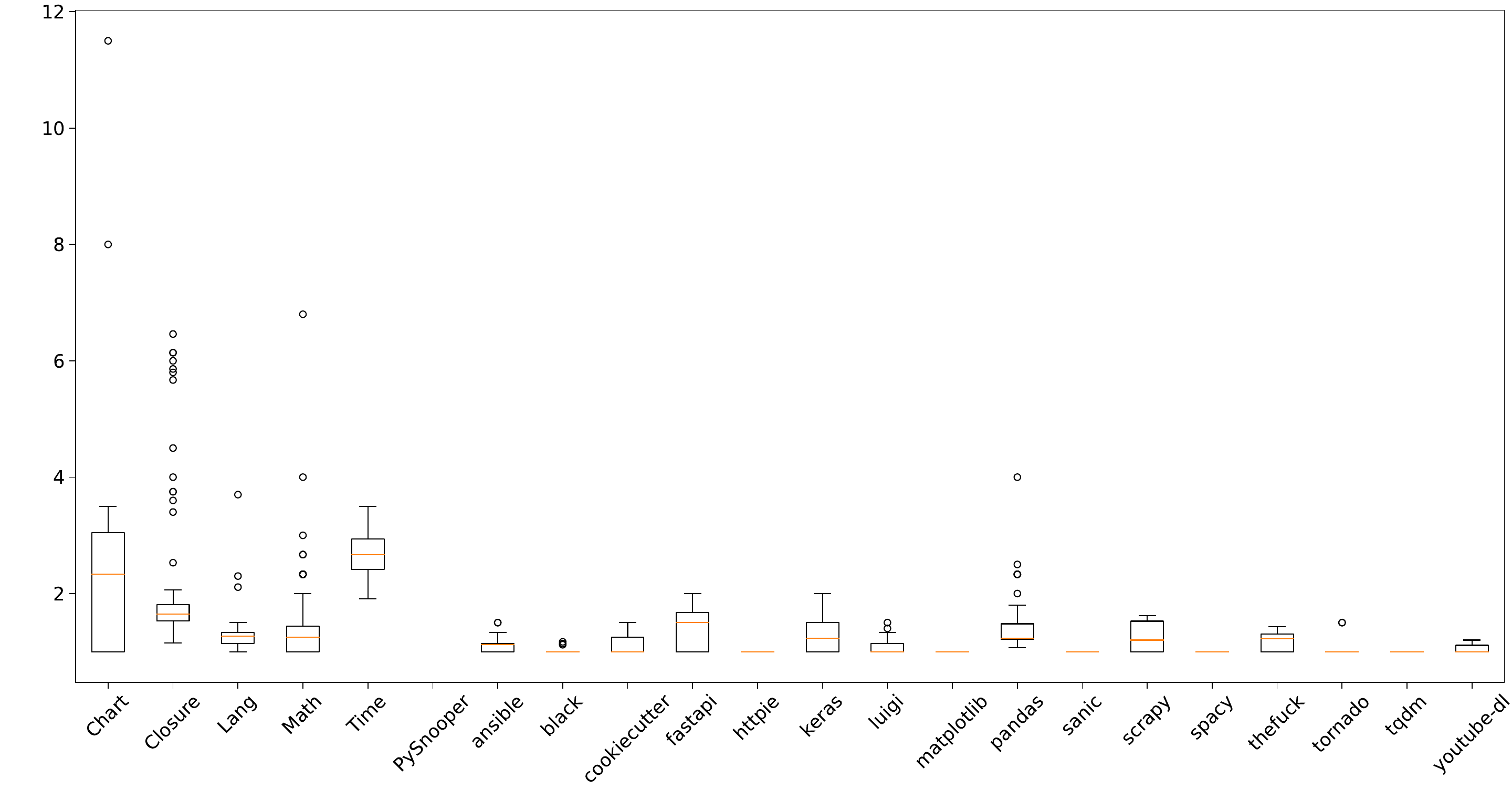}
  \caption{Number of tests transplanted per bug, averaged by version.}
  \label{fig:tpv}
\end{figure*}

This paper describes two separate datasets, Defects4J-mf~\cite{d4j-multi} and
BugsInPy-mf~\cite{bugsinpy-multi}, which we created using the same techniques and
for the same purposes. Both are multi-fault extensions to the original, underlying
datasets Defects4J and BugsInPy, respectively.

Similar to the original datasets (see~\Cref{sec:orig}), the dataset extensions
created in this paper consist of pre-existing, unaltered versions from
underlying open-source repositories maintained using version control software.
In addition to this, we too identify existing bugs in the versions by means of
test case failures. However we differ from the original datasets by identifying
\emph{multiple faults} in each version. We do so by \emph{exposing} additional
faults in each version by transplanting test cases committed in future
versions (see~\Cref{sec:test-transplant}), and by \emph{identifying} the faulty
code locations by translating the fault locations identified by the original
datasets for the previous versions (see~\Cref{sec:location-translate}).
In order to ensure correctness, we only consider a bug as existing in
a version if the test case transplantation and fault location translation
processes both succeed; That is, if the bug is both \emph{exposed} in the
version by a failing test case, and \emph{identifiable}
by at least one line of code.

We successfully identify 9.2 respectively 18.6 faults in each version
from the Defecst4J respectively BugsInPy datasets.
\Cref{tab:stats} gives the overall dataset statistics, while
\Cref{fig:bug_stats} gives a more detailed look at the bug distributions.
We see from this figure that the Defects4J versions have on average substantially
fewer bugs (normalized by program size) identified in our datasets than the BugsInPy
versions, and that particular projects within BugsInPy have substantially higher
bug densities in their versions than the rest.
For each of these versions, we transplant on average 16.5 and 6.3 test cases for
Defects4J and BugsInPy respectively, which are necessary to expose the additional
bugs in these versions. \Cref{fig:tpv} shows the number of tests transplanted per
bug in each version.
We also report in~\Cref{tab:stats} the number of times a fault was excluded from
a version (drop rate), with the test case transplantation process succeeding, but
the fault location translation process failing. This indicates the number of times
a fault is exposed, but cannot be automatically identified in the version. On average,
this occurs 0.3\% and 14.3\% for Defects4J and BugsInPy respectively, with the
anomaly occurring more frequently on certain projects.

\begin{figure*}[ht]
  \centering
  \includegraphics[width=\textwidth]{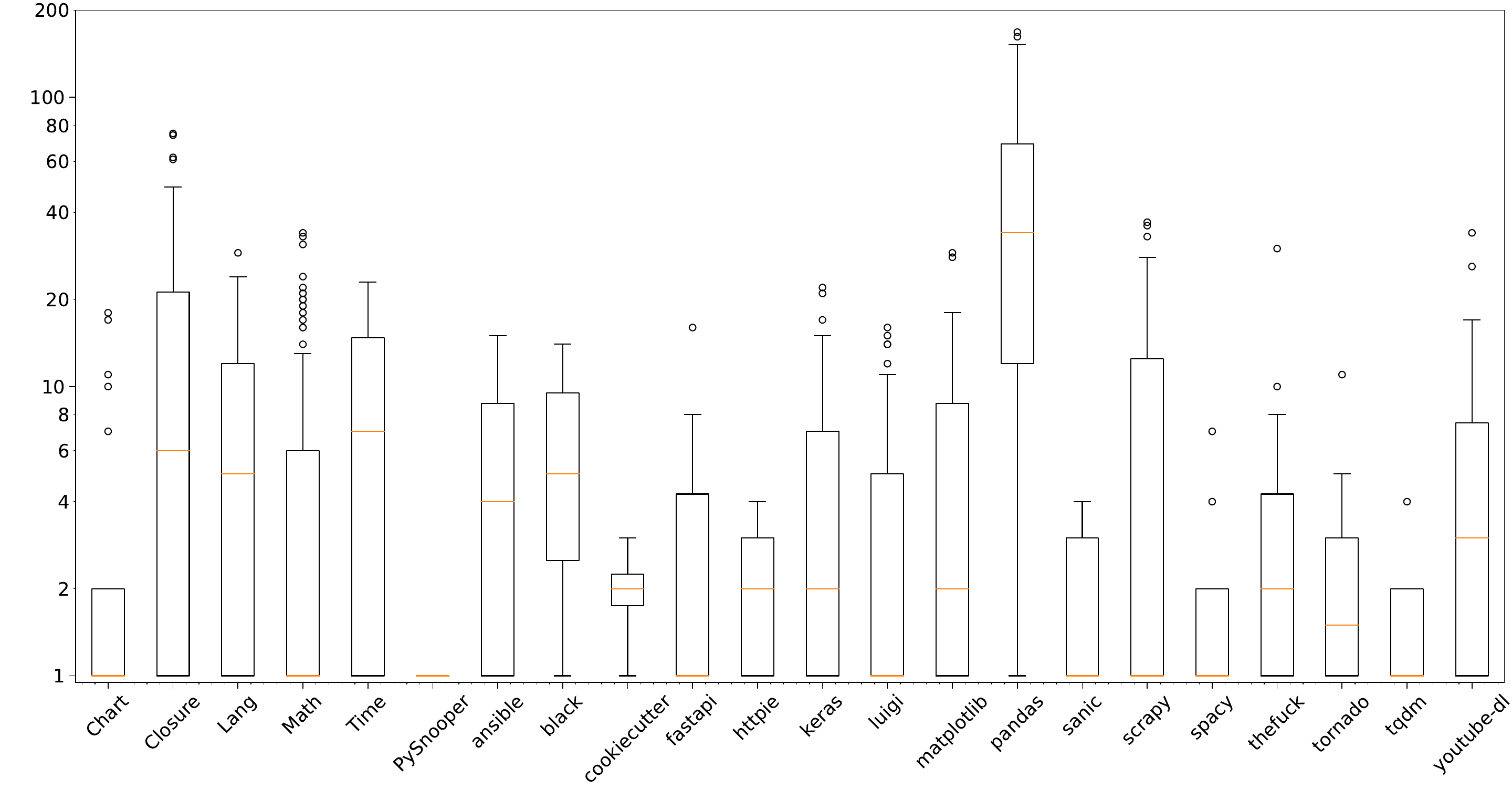}
  \caption{Average number of versions a particular bug is available in (y-axis
  in log scale).}
  \label{fig:bug_lifetimes}
\end{figure*}

\begin{figure*}[ht]
  \centering
  \includegraphics[width=\textwidth]{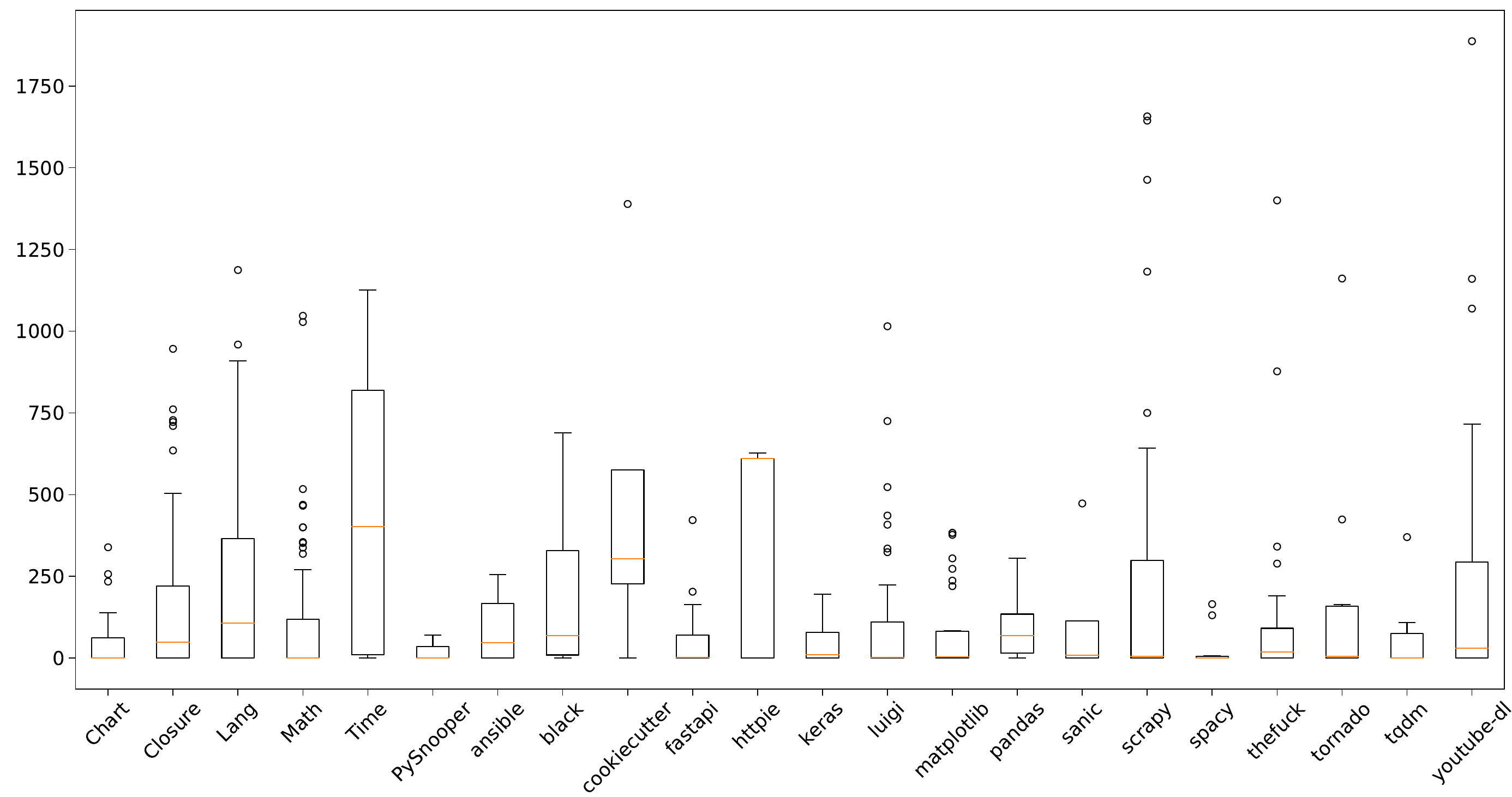}
  \caption{Average number of days between the oldest version a bug is available
  in and the version in which the bug is fixed (bug lifetime).}
  \label{fig:bug_lifetimes_times}
\end{figure*}


The datasets created in this paper also enable detailed perspectives on the
underlying software projects and versions themselves. \Cref{fig:bug_lifetimes}
gives one such insight, showing the average number of versions in which a particular
bug is available from each of the projects. Combining this information with the
information from each projects' git history, we are able to estimate the amount
of time a particular bug is active for, which is given
in~\Cref{fig:bug_lifetimes_times}. These figures give an estimate of the
average lifespan of a bug in a particular program. We note, however, that this
is a lower estimate on the lifespan of the bugs, as these bugs could be
available in more versions that are not identified by our techniques. As we can
see from~\Cref{fig:bug_lifetimes}, bugs from the Java-based Defects4J projects
last on average 6.9 Defects4J versions, whereas bugs from the Python-based BugsInPy
projects last only 4.1 BugsInPy versions on average (excluding the project
pandas). Despite this, there are particular bugs and whole projects (such as pandas)
where the average version lifespan is much greater. For example, most projects have
at least one bug that has a lifespan of on average 35 versions.
\Cref{fig:bug_lifetimes_times} indicates that the average lifespan of a bug is usually
quite small (around one to two weeks), however we also see here that bug lifetimes
vary widely, and that most projects again have at least one bug that lasts more
than 100 to 200 days). These statistics indicate that although it is uncommon for
bugs to last more than a week or two, there are usually individual bugs whose lifespan
spans a larger portion of the project history. These findings are corroborated in
current literature on the topic of bug lifetimes~\cite{bugLifetimes1, bugLifetimes2,
bugLifetimes3}, which indicates both the veracity of the data, and the accuracy
of our datasets in identifying faults within versions.

\section{Dataset usage}
As described in~\Cref{sec:background}, Defects4J and BugsInPy consist of
versions from popular open source projects written in Java and Python, however
the datasets themselves do not store each version for the projects, but
rather provide the facilities to easily clone the versions tracked by the dataset
from the original project repositories. We maintain the functionality and
setup of the original datasets in our extension,
and only add functionality to
allow each version to be identified as containing multiple faults.

\subsection{Usage description of the original datasets}\label{sec:origUsage}
For completeness, we describe the usage of the original
Defects4J and BugsInPy datasets.
Both allow interaction with the underlying
project versions through the use of a list of specialized
CLI commands. \Cref{tab:commands} lists the commands supported by both
Defects4J and BugsInPy.
They are run as \texttt{defects4j~<command>} and
\texttt{bugsinpy-<command>}, respectively.
%
Note that any of the provided tools can be run on
the multi-fault datasets described in this paper.


\begin{table}\setlength{\tabcolsep}{3pt}
  \centering
  \begin{tabular}{|l|l|}
    \hline
    \textbf{Command} & \textbf{Description}\\
    \hline
    \texttt{info} & Get the information of a specific project or bug\\
    \hline
    \textbf{\texttt{checkout}} & Checkout a buggy or a fixed project version\\
      &(use \textbf{\texttt{multi-checkout}} for BugsInPy multi-fault)\\
    \hline
    \texttt{compile} & Compile sources and developer-written tests of\\&a buggy or a fixed project version\\
    \hline
    \texttt{test} & Run a single test method or a test suite on a\\&buggy or a fixed project version\\
    \hline
    \textbf{\texttt{coverage}} & Run code coverage analysis on a buggy or a\\&fixed project version\\
    \hline
    \textbf{\texttt{to-tcm}} & Output coverage in TCM format (BugsInPy only)\\
    \hline
    \texttt{mutation} & Run mutation analysis on a buggy or a fixed\\&project version\\
    \hline
    \texttt{fuzz} & Run a test input generation from specific bug\\&(BugsInPy only)\\
    \hline
    \textbf{\texttt{identify}} & Add fault location information to elements\\
    \hline
  \end{tabular}
	\caption{Defects4J and BugsInPy commands; multi-fault modifications and extensions in bold.}
  \label{tab:commands}
\end{table}

\subsection{Usage description of the multi-fault datasets}
The main addition with our multi-fault datasets
is the
ability to identify versions in the underlying datasets that contain multiple
faults. As such, the main difference in the commands is the addition of a
multi-fault \texttt{checkout} command.
These can be run as \texttt{defects4j\_multi checkout} and
\texttt{bugsinpy-multi-checkout}, respectively. These commands
use the underlying Defects4j and BugsInPy datasets' \texttt{checkout}
commands to clone the version from the project repository; however, they
also add for \emph{each} of the faults identified in the version
the fault-exposing tests and the fault locations.
In both
multi-fault datasets, the
transplanted test cases are added to the existing test suite by a process of
test case source code alteration (``splicing''). After the multi-fault checkout
process is complete, these test cases are accessible in the test suite, and for
any test suite related commands in the underlying dataset. The fault locations
are available in \texttt{bug.locations.<bugId>} files for each bug, in both
datasets.

In addition to the \texttt{checkout} command, the our datasets
also provide useful commands for evaluation purposes. In particular, the
coverage commands provided by the original Defects4J and BugsInPy datasets do
not collect code coverage per test case which is needed for techniques such as
spectrum-based fault localization~\cite{eval}. We thus additionally provide
commands for this purpose. For Defects4J, we alter the original \texttt{coverage}
command to collect code coverage using Gzoltar~\cite{gzoltar} instead of
Jacoco~\cite{jacoco}. For BugsInPy, we change the settings of the
\texttt{coverage} command (which uses Python's \texttt{coverage.py} coverage library) to
extract coverage per test, and provide the \texttt{to-tcm} command, which
converts the collected coverage into TCM~\cite{tcm} format. We then also provide
commands in both datasets for identifying each of the faults, based on the fault
identification information, within the collected coverage format, using
the \texttt{identify} command. For both datasets, this
command adds the fault as a part of the element (i.e., source code line) name, in
the respective format.

\section{Dataset creation}
Like the original datasets, our multi-fault versions guarantee fault exposure
and fault location identification. The former is achieved
by test case transplantation, the latter
by fault location translation.
We describe each step in turn.

\subsection{Test case transplantation}\label{sec:test-transplant}
Test case transplantation copies fault-revealing test cases from the
test suite of one bug repository entry to that of an earlier entry.
This process does not alter the source code of the project's versions, and all
test case logic is \emph{extracted} from an existing projcet version, and not
\emph{created}. The top of \Cref{fig:big_picture} shows the test case
transplantation process \circled{7}: test cases which expose the fault in the
buggy version of an entry are extracted, and then copied to a previous entry.
The test cases are then compiled and run, and their
output is compared with their output from their original version. If the
outputs are similar enough according to the Hunt-Szymanski
algorithm~\cite{HuntSzymanski} for longest common subsequence (LCS), the fault
is considered exposed also in the target entry.
Each set of fault-exposing test cases is transplanted as far as possible, i.e., until the fault is no longer exposed.


For the Defects4J, we reused the the test case transplantation by
An et.\ al.~\cite{multid4j}. They provide tools for extracting and copying the
test cases from one version to another, and identify many Defects4J versions in
which test cases can be transplanted to expose multiple faults. Their tools and
results have been included in the defects4j-mf dataset created in this
paper. We note that An et.\ al.\ were only able to identify 311 multi-fault
versions out of the 396 available bugs in Defects4j v1.0.1. We thus only include
these 311 versions in our defects4j-mf dataset.

For BugsInPy, we carried out a similar process in order to achieve the same
results. In particular, we provide the tools for extracting and copying test
cases from one version to another, and identify the BugsInPy versions in which
test cases can be transplanted for the exposure of multiple faults. This
process was considerably more complex for the Python project versions in
BugsInPy, due to certain Python coding conventions, which encourage test
fragment reuse, and specialized imports. We developed a \emph{source code
dependency aware} test extraction and copying tool, which allows both the test
cases and their respective source code dependencies (e.g.\ test fixtures,
imports, etc.) to be extracted and copied between versions.

\subsection{Fault location translation}\label{sec:location-translate}
In the original Defects4J and BugsInPy datasets, the fault identification used
for the location oracle for each buggy version $b$ either uses the lines
changed in the diff $\Delta_f$ as an approximation \circled{4}, or relies on a
more precise manual identification \circled{5}.
We use either of these methods as a starting point for fault identification in
the multi-fault versions.  However, these identified locations cannot be used
directly to identify the fault locations in prior versions as the changes
during the development may have caused the source code locations to shift.
%
%
We therefore backtrack~\cite{bug-backtracker} the starting locations through all versions in the
complete project repository, until we reach the buggy target version $b'$ of
each test case transplantation step; for each version $i$, we consider the
operations in the diff $\Delta_i$ and update the fault locations as follows
\circled{8}.
\begin{enumerate*}
\item If a source file containing a tracked location is
    renamed, the tracking respects this renaming.
\item If source code in the same file as a tracked location is
    altered above this location, then the tracked location is adjusted to
    reflect the changes.
\item If a tracked location is modified or added in a particular
    diff, then tracking for this location is stopped; this
    ensures that the tracked source lines remain unmodified, and are thus
    identical to the location in the version in which the bug was originally
    identified.
\end{enumerate*}
We consider a particular fault identified in a target version if at least one
identified fault location is tracked successfully back into that version.
\subsection{Limitations and threats to validity}\label{sec:threats}
The identification of buggy versions from the underlying real-world, open-source
projects used in the Defects4J and BugsInPy datasets was done manually.
This manual identification of buggy versions is occasionally incorrect and can
lead to incorrect results in the multi-fault versions of the datasets.

Due to the extensive manual effort
we did not manually identify the location of each fault in each multi-fault
version. Where available, we used existing manual fault
identification~\cite{d4j_bug_identify} for each bug in the version where it was
discovered by Defects4J or BugsInPy; otherwise, we used the
diff $\Delta_f$ between the buggy and fixed versions as an approximation. We
then extended this to the multi-fault versions using the automated fault
location translation process.
The approximate fault identification using source code diffs and the automated
fault location translation are both susceptible to errors; in particular the
location translation may be unable to trace any faulty lines to an earlier
version and thus fail to identify all faults. We automatically verified that
lines were properly translated, and manually corrected lines incorrectly
translated.

The test case transplantation process may also not always
work correctly. In order to prevent this from interfering with the quality of
the dataset, we tested each transplanted test case, and only accepted
transplanted test cases that compiled (i.e., did not produce any compile-time
or runtime errors) and produced the same result (i.e., expected output or
error) as in the original Defects4J or BugsInPy version, and whose fault
location could be fully translated. This ensures a conservative approach; only
bugs that are truly exposed in a version are identified, but some bugs that may
actually be active in a version may be missed.

As noted in~\cite{nofuture}, the underlying datasets used in this paper may have
the limitation that their test cases are usually only available in the project
version which contains the corresponding bug, and thus could be contaminated by
the knowledge of the bug-fix. This limitation may also therefore have an impact
on the datasets created in this paper. In addition, this limitation may be compounded
by the fact that the test case transplantation process used in this paper modifies
the test suite of a version by including test cases that were only available in
subsequent versions, and thus could contaminate previous versions with the
knowledge of fixes from later versions. This limitation may have an adverse effect,
mainly on automated program repair techniques, as discussed in~\cite{nofuture},
but may also have a minor effect on localization techniques. However based on a
sample of bugs from each underlying dataset, we notice that the test cases can
be reasonably constructed from the corresponding bug report without the knowledge
of the bug-fix, indicating a lack of dependence of the test cases on the future
knowledge. This suggests that this limitation may be mitigated by further study
on the composition of the underlying datasets' test suites. We leave such study
as future work.

The fault location translation process used in this paper to identify the
faulty locations does not allow the inclusion of the developer-written bug patches
in the multi-fault versions. This presents a limitation for techniques such as
automated program repair (APR), which often require these patches for evaluation
of the techniques.

\section{Future Dataset Usage}

\subsection{Multi-fault localization}
Multi-fault localization \cite{DBLP:conf/kbse/AbreuZG09} and program repair are
open problems; multi-fault datasets mined from real-world software projects such
as these described here will therefore be useful for training and evaluation of
multi-fault debugging tools.
More specifically, we already used our multi-fault version of Defects4J for the
evaluation of our spectrum-based fault localization tool FLITSR~\cite{flitsr},
and showed that it can localize multiple faults at the same time.


Automated debugging techniques that rely on machine learning also require large
datasets of bugs as training data. Previously, these techniques have used
datasets of synthetic (injected) faults, and single fault datasets such as
Defects4J~\cite{grace,deepFl}. However, this leads to bias in the machine
learning model~\cite{flitsr}.  Our multi-fault datasets can be used as more
realistic training data for such machine learning models to improve real-world
applicability.

We also see qualitative uses of our datasets. In particular, the identification
of software project versions with multiple bugs existing simultaneously may be
used in an analysis of
the presence of multiple bugs in software systems.
In addition, the fact that test cases could be transplanted from newer versions
to expose bugs in previous versions may also provide insight into research
questions such as ``can better test suites expose more bugs?''.
This has applications in software fuzzing and automated test case generation.

\subsection{Future work}
Due to automated data construction, any improvements
of the underlying Defects4J and BugsInPy datasets will improve the quality
of our multi-fault versions as well. This leads to many avenues for further
improvement of this dataset by improving:
\begin{enumerate*}
  \item the fault identification for each bug through manual fault
    identification as in~\cite{d4j_bug_identify} for other projects,
  \item the bug isolation and reproducibility by better automation of the set-up
    for each version, and
  \item the size of the datasets by adding more versions exposing more bugs.
\end{enumerate*}
Each of the above changes will result in improvements in the multi-fault
counterparts as well, allowing for better identification, reproducibility,
and more bugs in each multi-fault version.


The fault location translation process does not yet fully support complex branching
in the git history. We leave it as future work to add this functionality.
The ideas used to extend the Defects4J and BugsInPy datasets in this paper can
be generalized to other datasets involving many other languages. An example of
such a dataset to which these techniques can be applied is the HasBugs
dataset~\cite{hasbugs}.
We leave the
extension of such datasets using the techniques provided as future work.
As mentioned in~\Cref{sec:threats}, bug patches for all faults are not included
in the multi-fault versions. We identify the addition of these patches as future
work.

\section{Conclusion}
In this paper we present extensions to the Defects4J and BugsInPy datasets of
bugs in real-world software projects which expose the existence of multiple
bugs in each of the versions. We find on average 9.2 and 18.6 bugs in
the Defects4J and BugsInPy versions, respectively. The extension uses
test case transplantation and fault location translation to
identify these multi-fault versions. In doing so, we do not create or modify any
of the existing real-world software project's code, and only use test cases
produced by the developers of the corresponding software project.

We have made the creation of the multi-fault extension of the datasets
mostly automatic, simplifying reproducibility and future verification. In addition,
the process requires only minimal manual efforts when run on new projects or versions
added to the underlying Defects4J and BugsInPy datasets.

We maintain the existing frameworks' extensibility and ease of use by allowing
all existing functionality to be used in the extensions. We additionally add useful
functionality for coverage collection for use in fault localization and program
repair.

\bibliographystyle{ACM-Reference-Format}
\bibliography{refs}


\begin{thebibliography}{47}


\ifx \showCODEN    \undefined \def \showCODEN     #1{\unskip}     \fi
\ifx \showDOI      \undefined \def \showDOI       #1{#1}\fi
\ifx \showISBNx    \undefined \def \showISBNx     #1{\unskip}     \fi
\ifx \showISBNxiii \undefined \def \showISBNxiii  #1{\unskip}     \fi
\ifx \showISSN     \undefined \def \showISSN      #1{\unskip}     \fi
\ifx \showLCCN     \undefined \def \showLCCN      #1{\unskip}     \fi
\ifx \shownote     \undefined \def \shownote      #1{#1}          \fi
\ifx \showarticletitle \undefined \def \showarticletitle #1{#1}   \fi
\ifx \showURL      \undefined \def \showURL       {\relax}        \fi
\providecommand\bibfield[2]{#2}
\providecommand\bibinfo[2]{#2}
\providecommand\natexlab[1]{#1}
\providecommand\showeprint[2][]{arXiv:#2}

\bibitem[\protect\citeauthoryear{??}{tcm}{2014}]%
        {tcm}
 \bibinfo{year}{2014}\natexlab{}.
\newblock \bibinfo{title}{{``More Debugging in Parallel'' Resource Page}}.
\newblock
  \bibinfo{howpublished}{\url{https://www.fernuni-hagen.de/ps/prjs/PD/}}.
\newblock


\bibitem[\protect\citeauthoryear{??}{jac}{2017}]%
        {jacoco}
 \bibinfo{year}{2017}\natexlab{}.
\newblock \bibinfo{title}{{JaCoCo Java Code Coverage Library}}.
\newblock \bibinfo{howpublished}{\url{https://www.eclemma.org/jacoco/}}.
\newblock


\bibitem[\protect\citeauthoryear{??}{bug}{2023a}]%
        {bugsinpy-multi}
 \bibinfo{year}{2023}\natexlab{a}.
\newblock \bibinfo{title}{{BugsInPy multi-fault repository}}.
\newblock \bibinfo{howpublished}{\url{https://github.com/DCallaz/bugsinpy-mf}}.
\newblock


\bibitem[\protect\citeauthoryear{??}{d4j}{2023}]%
        {d4j-multi}
 \bibinfo{year}{2023}\natexlab{}.
\newblock \bibinfo{title}{{Defects4J multi-fault repository}}.
\newblock
  \bibinfo{howpublished}{\url{https://github.com/DCallaz/defects4j-mf}}.
\newblock


\bibitem[\protect\citeauthoryear{??}{bug}{2023b}]%
        {bug-backtracker}
 \bibinfo{year}{2023}\natexlab{b}.
\newblock \bibinfo{title}{{Fault location translation (backtracking) tool}}.
\newblock
  \bibinfo{howpublished}{\url{https://github.com/DCallaz/bug-backtracker}}.
\newblock


\bibitem[\protect\citeauthoryear{Abreu, Zoeteweij, and van Gemund}{Abreu
  et~al\mbox{.}}{2009}]%
        {DBLP:conf/kbse/AbreuZG09}
\bibfield{author}{\bibinfo{person}{Rui Abreu}, \bibinfo{person}{Peter
  Zoeteweij}, {and} \bibinfo{person}{Arjan J.~C. van Gemund}.}
  \bibinfo{year}{2009}\natexlab{}.
\newblock \showarticletitle{Spectrum-Based Multiple Fault Localization}. In
  \bibinfo{booktitle}{\emph{{ASE} 2009, 24th {IEEE/ACM} International
  Conference on Automated Software Engineering, Auckland, New Zealand, November
  16-20, 2009}}. \bibinfo{publisher}{{IEEE} Computer Society},
  \bibinfo{pages}{88--99}.
\newblock
\urldef\tempurl%
\url{https://doi.org/10.1109/ASE.2009.25}
\showDOI{\tempurl}


\bibitem[\protect\citeauthoryear{An, Yoon, and Yoo}{An et~al\mbox{.}}{2021}]%
        {multid4j}
\bibfield{author}{\bibinfo{person}{Gabin An}, \bibinfo{person}{Juyeon Yoon},
  {and} \bibinfo{person}{Shin Yoo}.} \bibinfo{year}{2021}\natexlab{}.
\newblock \showarticletitle{Searching for Multi-fault Programs in {Defects4J}}.
  In \bibinfo{booktitle}{\emph{Search-Based Software Engineering - 13th
  International Symposium, {SSBSE} 2021, Bari, Italy, October 11-12, 2021,
  Proceedings}} \emph{(\bibinfo{series}{Lecture Notes in Computer Science})},
  Vol.~\bibinfo{volume}{12914}. \bibinfo{publisher}{Springer},
  \bibinfo{pages}{153--158}.
\newblock
\urldef\tempurl%
\url{https://doi.org/10.1007/978-3-030-88106-1\_11}
\showDOI{\tempurl}


\bibitem[\protect\citeauthoryear{{Apache Software Foundation}}{{Apache Software
  Foundation}}{2002}]%
        {lang}
\bibfield{author}{\bibinfo{person}{{Apache Software Foundation}}.}
  \bibinfo{year}{2002}\natexlab{}.
\newblock \bibinfo{title}{Commons Lang}.
\newblock
\newblock
\urldef\tempurl%
\url{https://commons.apache.org/proper/commons-lang/}
\showURL{%
\tempurl}


\bibitem[\protect\citeauthoryear{{Apache Software Foundation}}{{Apache Software
  Foundation}}{2007}]%
        {math}
\bibfield{author}{\bibinfo{person}{{Apache Software Foundation}}.}
  \bibinfo{year}{2007}\natexlab{}.
\newblock \bibinfo{title}{Apache Commons Math}.
\newblock
\newblock
\urldef\tempurl%
\url{https://commons.apache.org/proper/commons-math/}
\showURL{%
\tempurl}


\bibitem[\protect\citeauthoryear{Applis and Panichella}{Applis and
  Panichella}{2023}]%
        {hasbugs}
\bibfield{author}{\bibinfo{person}{Leonhard Applis} {and}
  \bibinfo{person}{Annibale Panichella}.} \bibinfo{year}{2023}\natexlab{}.
\newblock \showarticletitle{HasBugs - Handpicked Haskell Bugs}. In
  \bibinfo{booktitle}{\emph{20th {IEEE/ACM} International Conference on Mining
  Software Repositories, {MSR} 2023, Melbourne, Australia, May 15-16, 2023}}.
  \bibinfo{publisher}{{IEEE}}, \bibinfo{pages}{223--227}.
\newblock
\urldef\tempurl%
\url{https://doi.org/10.1109/MSR59073.2023.00040}
\showDOI{\tempurl}


\bibitem[\protect\citeauthoryear{Bernhardsson, Freider, and {contributors to
  Luigi}}{Bernhardsson et~al\mbox{.}}{2012}]%
        {luigi}
\bibfield{author}{\bibinfo{person}{Erik Bernhardsson}, \bibinfo{person}{Elias
  Freider}, {and} \bibinfo{person}{{contributors to Luigi}}.}
  \bibinfo{year}{2012}\natexlab{}.
\newblock \bibinfo{title}{Luigi}.
\newblock
\newblock
\urldef\tempurl%
\url{https://github.com/spotify/luigi}
\showURL{%
\tempurl}


\bibitem[\protect\citeauthoryear{Bolton and {contributors to
  Youtube-dl}}{Bolton and {contributors to Youtube-dl}}{2011}]%
        {youtube-dl}
\bibfield{author}{\bibinfo{person}{Daniel Bolton} {and}
  \bibinfo{person}{{contributors to Youtube-dl}}.}
  \bibinfo{year}{2011}\natexlab{}.
\newblock \bibinfo{title}{Youtube-dl}.
\newblock
\newblock
\urldef\tempurl%
\url{https://github.com/ytdl-org/youtube-dl}
\showURL{%
\tempurl}


\bibitem[\protect\citeauthoryear{Callaghan and Fischer}{Callaghan and
  Fischer}{2023}]%
        {flitsr}
\bibfield{author}{\bibinfo{person}{Dylan Callaghan} {and}
  \bibinfo{person}{Bernd Fischer}.} \bibinfo{year}{2023}\natexlab{}.
\newblock \showarticletitle{Improving Spectrum-Based Localization of Multiple
  Faults by Iterative Test Suite Reduction}. In
  \bibinfo{booktitle}{\emph{Proceedings of the 32nd {ACM} {SIGSOFT}
  International Symposium on Software Testing and Analysis, {ISSTA} 2023,
  Seattle, WA, USA, July 17-21, 2023}},
  \bibfield{editor}{\bibinfo{person}{Ren{\'{e}} Just} {and}
  \bibinfo{person}{Gordon Fraser}} (Eds.). \bibinfo{publisher}{{ACM}},
  \bibinfo{pages}{1445--1457}.
\newblock
\urldef\tempurl%
\url{https://doi.org/10.1145/3597926.3598148}
\showDOI{\tempurl}


\bibitem[\protect\citeauthoryear{Campos, Riboira, Perez, and Abreu}{Campos
  et~al\mbox{.}}{2012}]%
        {gzoltar}
\bibfield{author}{\bibinfo{person}{Jos{\'{e}} Campos},
  \bibinfo{person}{Andr{\'{e}} Riboira}, \bibinfo{person}{Alexandre Perez},
  {and} \bibinfo{person}{Rui Abreu}.} \bibinfo{year}{2012}\natexlab{}.
\newblock \showarticletitle{{GZoltar}: an eclipse plug-in for testing and
  debugging}. In \bibinfo{booktitle}{\emph{{IEEE/ACM} International Conference
  on Automated Software Engineering, ASE'12, Essen, Germany, September 3-7,
  2012}}. \bibinfo{publisher}{{ACM}}, \bibinfo{pages}{378--381}.
\newblock
\urldef\tempurl%
\url{https://doi.org/10.1145/2351676.2351752}
\showDOI{\tempurl}


\bibitem[\protect\citeauthoryear{Canfora, Ceccarelli, Cerulo, and
  Penta}{Canfora et~al\mbox{.}}{2011}]%
        {bugLifetimes1}
\bibfield{author}{\bibinfo{person}{Gerardo Canfora}, \bibinfo{person}{Michele
  Ceccarelli}, \bibinfo{person}{Luigi Cerulo}, {and}
  \bibinfo{person}{Massimiliano~Di Penta}.} \bibinfo{year}{2011}\natexlab{}.
\newblock \showarticletitle{How Long Does a Bug Survive? An Empirical Study}.
  In \bibinfo{booktitle}{\emph{18th Working Conference on Reverse Engineering,
  {WCRE} 2011, Limerick, Ireland, October 17-20, 2011}},
  \bibfield{editor}{\bibinfo{person}{Martin Pinzger}, \bibinfo{person}{Denys
  Poshyvanyk}, {and} \bibinfo{person}{Jim Buckley}} (Eds.).
  \bibinfo{publisher}{{IEEE} Computer Society}, \bibinfo{pages}{191--200}.
\newblock
\urldef\tempurl%
\url{https://doi.org/10.1109/WCRE.2011.31}
\showDOI{\tempurl}


\bibitem[\protect\citeauthoryear{Chollet et~al\mbox{.}}{Chollet
  et~al\mbox{.}}{2015}]%
        {keras}
\bibfield{author}{\bibinfo{person}{Fran\c{c}ois Chollet} {et~al\mbox{.}}}
  \bibinfo{year}{2015}\natexlab{}.
\newblock \bibinfo{title}{Keras}.
\newblock \bibinfo{howpublished}{\url{https://keras.io}}.
\newblock


\bibitem[\protect\citeauthoryear{{Closure Compiler Authors}}{{Closure Compiler
  Authors}}{2009}]%
        {closure}
\bibfield{author}{\bibinfo{person}{{Closure Compiler Authors}}.}
  \bibinfo{year}{2009}\natexlab{}.
\newblock \bibinfo{title}{Closure Compiler}.
\newblock
\newblock
\urldef\tempurl%
\url{https://developers.google.com/closure/compiler/}
\showURL{%
\tempurl}


\bibitem[\protect\citeauthoryear{Colebourne and {contributors to
  Joda-Time}}{Colebourne and {contributors to Joda-Time}}{2014}]%
        {time}
\bibfield{author}{\bibinfo{person}{Stephen Colebourne} {and}
  \bibinfo{person}{{contributors to Joda-Time}}.}
  \bibinfo{year}{2014}\natexlab{}.
\newblock \bibinfo{title}{Joda-Time}.
\newblock
\newblock
\urldef\tempurl%
\url{https://www.joda.org/joda-time/}
\showURL{%
\tempurl}


\bibitem[\protect\citeauthoryear{da~Costa-Luis}{da~Costa-Luis}{2019}]%
        {tqdm}
\bibfield{author}{\bibinfo{person}{Casper~O da Costa-Luis}.}
  \bibinfo{year}{2019}\natexlab{}.
\newblock \showarticletitle{tqdm: A fast, extensible progress meter for python
  and cli}.
\newblock \bibinfo{journal}{\emph{Journal of Open Source Software}}
  \bibinfo{volume}{4}, \bibinfo{number}{37} (\bibinfo{year}{2019}),
  \bibinfo{pages}{1277}.
\newblock


\bibitem[\protect\citeauthoryear{DeHaan and {contributors to Ansible}}{DeHaan
  and {contributors to Ansible}}{2013}]%
        {ansible}
\bibfield{author}{\bibinfo{person}{Michael DeHaan} {and}
  \bibinfo{person}{{contributors to Ansible}}.}
  \bibinfo{year}{2013}\natexlab{}.
\newblock \bibinfo{title}{Ansible}.
\newblock
\newblock
\urldef\tempurl%
\url{https://github.com/ansible/ansible}
\showURL{%
\tempurl}


\bibitem[\protect\citeauthoryear{Do, Elbaum, and Rothermel}{Do
  et~al\mbox{.}}{2005}]%
        {SIR}
\bibfield{author}{\bibinfo{person}{Hyunsook Do}, \bibinfo{person}{Sebastian~G.
  Elbaum}, {and} \bibinfo{person}{Gregg Rothermel}.}
  \bibinfo{year}{2005}\natexlab{}.
\newblock \showarticletitle{Supporting Controlled Experimentation with Testing
  Techniques: An Infrastructure and its Potential Impact}.
\newblock \bibinfo{journal}{\emph{Empir. Softw. Eng.}} \bibinfo{volume}{10},
  \bibinfo{number}{4} (\bibinfo{year}{2005}), \bibinfo{pages}{405--435}.
\newblock
\urldef\tempurl%
\url{https://doi.org/10.1007/S10664-005-3861-2}
\showDOI{\tempurl}


\bibitem[\protect\citeauthoryear{Gilbert and {contributors to
  JFreeChart}}{Gilbert and {contributors to JFreeChart}}{2000}]%
        {chart}
\bibfield{author}{\bibinfo{person}{David Gilbert} {and}
  \bibinfo{person}{{contributors to JFreeChart}}.}
  \bibinfo{year}{2000}\natexlab{}.
\newblock \bibinfo{title}{JFreeChart}.
\newblock
\newblock
\urldef\tempurl%
\url{https://www.jfree.org/jfreechart}
\showURL{%
\tempurl}


\bibitem[\protect\citeauthoryear{Greenfeld and {contributors to
  Cookiecutter}}{Greenfeld and {contributors to Cookiecutter}}{2014}]%
        {cookiecutter}
\bibfield{author}{\bibinfo{person}{Audrey~Roy Greenfeld} {and}
  \bibinfo{person}{{contributors to Cookiecutter}}.}
  \bibinfo{year}{2014}\natexlab{}.
\newblock \bibinfo{title}{Cookiecutter}.
\newblock
\newblock
\urldef\tempurl%
\url{https://github.com/cookiecutter/cookiecutter}
\showURL{%
\tempurl}


\bibitem[\protect\citeauthoryear{Heiden, Grunske, Kehrer, Keller, van Hoorn,
  Filieri, and Lo}{Heiden et~al\mbox{.}}{2019}]%
        {eval}
\bibfield{author}{\bibinfo{person}{Simon Heiden}, \bibinfo{person}{Lars
  Grunske}, \bibinfo{person}{Timo Kehrer}, \bibinfo{person}{Fabian Keller},
  \bibinfo{person}{Andr{\'{e}} van Hoorn}, \bibinfo{person}{Antonio Filieri},
  {and} \bibinfo{person}{David Lo}.} \bibinfo{year}{2019}\natexlab{}.
\newblock \showarticletitle{An evaluation of pure spectrum-based fault
  localization techniques for large-scale software systems}.
\newblock \bibinfo{journal}{\emph{Softw. Pract. Exp.}} \bibinfo{volume}{49},
  \bibinfo{number}{8} (\bibinfo{year}{2019}), \bibinfo{pages}{1197--1224}.
\newblock
\urldef\tempurl%
\url{https://doi.org/10.1002/spe.2703}
\showDOI{\tempurl}


\bibitem[\protect\citeauthoryear{Hogerle, Steimann, and Frenkel}{Hogerle
  et~al\mbox{.}}{2014}]%
        {parallel:steimann}
\bibfield{author}{\bibinfo{person}{Wolfgang Hogerle},
  \bibinfo{person}{Friedrich Steimann}, {and} \bibinfo{person}{Marcus
  Frenkel}.} \bibinfo{year}{2014}\natexlab{}.
\newblock \showarticletitle{More Debugging in Parallel}. In
  \bibinfo{booktitle}{\emph{25th {IEEE} International Symposium on Software
  Reliability Engineering, {ISSRE} 2014, Naples, Italy, November 3-6, 2014}}.
  \bibinfo{publisher}{{IEEE} Computer Society}, \bibinfo{pages}{133--143}.
\newblock
\urldef\tempurl%
\url{https://doi.org/10.1109/ISSRE.2014.29}
\showDOI{\tempurl}


\bibitem[\protect\citeauthoryear{Honnibal, Montani, Van~Landeghem, and
  Boyd}{Honnibal et~al\mbox{.}}{2020}]%
        {spacy}
\bibfield{author}{\bibinfo{person}{Matthew Honnibal}, \bibinfo{person}{Ines
  Montani}, \bibinfo{person}{Sofie Van~Landeghem}, {and}
  \bibinfo{person}{Adriane Boyd}.} \bibinfo{year}{2020}\natexlab{}.
\newblock \showarticletitle{{spaCy: Industrial-strength Natural Language
  Processing in Python}}.
\newblock  (\bibinfo{year}{2020}).
\newblock
\urldef\tempurl%
\url{https://doi.org/10.5281/zenodo.1212303}
\showDOI{\tempurl}


\bibitem[\protect\citeauthoryear{Hunt and Szymanski}{Hunt and
  Szymanski}{1977}]%
        {HuntSzymanski}
\bibfield{author}{\bibinfo{person}{James~W. Hunt} {and}
  \bibinfo{person}{Thomas~G. Szymanski}.} \bibinfo{year}{1977}\natexlab{}.
\newblock \showarticletitle{A Fast Algorithm for Computing Longest Common
  Subsequences}.
\newblock \bibinfo{journal}{\emph{Commun. ACM}} \bibinfo{volume}{20},
  \bibinfo{number}{5} (\bibinfo{date}{may} \bibinfo{year}{1977}),
  \bibinfo{pages}{350–353}.
\newblock
\showISSN{0001-0782}
\urldef\tempurl%
\url{https://doi.org/10.1145/359581.359603}
\showDOI{\tempurl}


\bibitem[\protect\citeauthoryear{Hunter}{Hunter}{2007}]%
        {matplotlib}
\bibfield{author}{\bibinfo{person}{John~D. Hunter}.}
  \bibinfo{year}{2007}\natexlab{}.
\newblock \showarticletitle{{Matplotlib: A 2D graphics environment}}.
\newblock \bibinfo{journal}{\emph{Computing in Science and Engineering}}
  \bibinfo{volume}{9}, \bibinfo{number}{3} (\bibinfo{year}{2007}),
  \bibinfo{pages}{90--95}.
\newblock
\urldef\tempurl%
\url{https://doi.org/10.1109/MCSE.2007.55}
\showDOI{\tempurl}


\bibitem[\protect\citeauthoryear{Iakovlev and {contributors to The
  Fuck}}{Iakovlev and {contributors to The Fuck}}{2015}]%
        {thefuck}
\bibfield{author}{\bibinfo{person}{Vladimir Iakovlev} {and}
  \bibinfo{person}{{contributors to The Fuck}}.}
  \bibinfo{year}{2015}\natexlab{}.
\newblock \bibinfo{title}{The Fuck}.
\newblock
\newblock
\urldef\tempurl%
\url{https://github.com/nvbn/thefuck}
\showURL{%
\tempurl}


\bibitem[\protect\citeauthoryear{Just, Jalali, and Ernst}{Just
  et~al\mbox{.}}{2014}]%
        {defects4j}
\bibfield{author}{\bibinfo{person}{Ren{\'{e}} Just}, \bibinfo{person}{Darioush
  Jalali}, {and} \bibinfo{person}{Michael~D. Ernst}.}
  \bibinfo{year}{2014}\natexlab{}.
\newblock \showarticletitle{D{efects4J}: a database of existing faults to
  enable controlled testing studies for Java programs}. In
  \bibinfo{booktitle}{\emph{International Symposium on Software Testing and
  Analysis, {ISSTA} '14, San Jose, CA, {USA} - July 21-26, 2014}}.
  \bibinfo{publisher}{{ACM}}, \bibinfo{pages}{437--440}.
\newblock
\urldef\tempurl%
\url{https://doi.org/10.1145/2610384.2628055}
\showDOI{\tempurl}


\bibitem[\protect\citeauthoryear{Kabadi, Kong, Xie, Bao, Prana, Le, Le, and
  Lo}{Kabadi et~al\mbox{.}}{2023}]%
        {nofuture}
\bibfield{author}{\bibinfo{person}{Vinay Kabadi}, \bibinfo{person}{Dezhen
  Kong}, \bibinfo{person}{Siyu Xie}, \bibinfo{person}{Lingfeng Bao},
  \bibinfo{person}{Gede Artha~Azriadi Prana}, \bibinfo{person}{Tien{-}Duy~B.
  Le}, \bibinfo{person}{Xuan{-}Bach~Dinh Le}, {and} \bibinfo{person}{David
  Lo}.} \bibinfo{year}{2023}\natexlab{}.
\newblock \showarticletitle{The Future Can't Help Fix The Past: Assessing
  Program Repair In The Wild}. In \bibinfo{booktitle}{\emph{{IEEE}
  International Conference on Software Maintenance and Evolution, {ICSME} 2023,
  Bogot{\'{a}}, Colombia, October 1-6, 2023}}. \bibinfo{publisher}{{IEEE}},
  \bibinfo{pages}{50--61}.
\newblock
\urldef\tempurl%
\url{https://doi.org/10.1109/ICSME58846.2023.00017}
\showDOI{\tempurl}


\bibitem[\protect\citeauthoryear{Kim and Jr.}{Kim and Jr.}{2006}]%
        {bugLifetimes3}
\bibfield{author}{\bibinfo{person}{Sunghun Kim} {and}
  \bibinfo{person}{E.~James~Whitehead Jr.}} \bibinfo{year}{2006}\natexlab{}.
\newblock \showarticletitle{How long did it take to fix bugs?}. In
  \bibinfo{booktitle}{\emph{Proceedings of the 2006 International Workshop on
  Mining Software Repositories, {MSR} 2006, Shanghai, China, May 22-23, 2006}},
  \bibfield{editor}{\bibinfo{person}{Stephan Diehl}, \bibinfo{person}{Harald~C.
  Gall}, {and} \bibinfo{person}{Ahmed~E. Hassan}} (Eds.).
  \bibinfo{publisher}{{ACM}}, \bibinfo{pages}{173--174}.
\newblock
\urldef\tempurl%
\url{https://doi.org/10.1145/1137983.1138027}
\showDOI{\tempurl}


\bibitem[\protect\citeauthoryear{Langa and {contributors to Black}}{Langa and
  {contributors to Black}}{2018}]%
        {black}
\bibfield{author}{\bibinfo{person}{Łukasz Langa} {and}
  \bibinfo{person}{{contributors to Black}}.} \bibinfo{year}{2018}\natexlab{}.
\newblock \bibinfo{booktitle}{\emph{{Black: The uncompromising Python code
  formatter}}}.
\newblock
\urldef\tempurl%
\url{https://github.com/psf/black}
\showURL{%
\tempurl}


\bibitem[\protect\citeauthoryear{Li, Li, Zhang, and Zhang}{Li
  et~al\mbox{.}}{2019}]%
        {deepFl}
\bibfield{author}{\bibinfo{person}{Xia Li}, \bibinfo{person}{Wei Li},
  \bibinfo{person}{Yuqun Zhang}, {and} \bibinfo{person}{Lingming Zhang}.}
  \bibinfo{year}{2019}\natexlab{}.
\newblock \showarticletitle{{DeepFL}: integrating multiple fault diagnosis
  dimensions for deep fault localization}. In
  \bibinfo{booktitle}{\emph{Proceedings of the 28th {ACM} {SIGSOFT}
  International Symposium on Software Testing and Analysis, {ISSTA} 2019,
  Beijing, China, July 15-19, 2019}}. \bibinfo{publisher}{{ACM}},
  \bibinfo{pages}{169--180}.
\newblock
\urldef\tempurl%
\url{https://doi.org/10.1145/3293882.3330574}
\showDOI{\tempurl}


\bibitem[\protect\citeauthoryear{Lou, Zhu, Dong, Li, Sun, Hao, Zhang, and
  Zhang}{Lou et~al\mbox{.}}{2021}]%
        {grace}
\bibfield{author}{\bibinfo{person}{Yiling Lou}, \bibinfo{person}{Qihao Zhu},
  \bibinfo{person}{Jinhao Dong}, \bibinfo{person}{Xia Li},
  \bibinfo{person}{Zeyu Sun}, \bibinfo{person}{Dan Hao}, \bibinfo{person}{Lu
  Zhang}, {and} \bibinfo{person}{Lingming Zhang}.}
  \bibinfo{year}{2021}\natexlab{}.
\newblock \showarticletitle{Boosting coverage-based fault localization via
  graph-based representation learning}. In \bibinfo{booktitle}{\emph{{ESEC/FSE}
  '21: 29th {ACM} Joint European Software Engineering Conference and Symposium
  on the Foundations of Software Engineering, Athens, Greece, August 23-28,
  2021}}. \bibinfo{publisher}{{ACM}}, \bibinfo{pages}{664--676}.
\newblock
\urldef\tempurl%
\url{https://doi.org/10.1145/3468264.3468580}
\showDOI{\tempurl}


\bibitem[\protect\citeauthoryear{Rachum, Hall, Yanokura, et~al\mbox{.}}{Rachum
  et~al\mbox{.}}{2019}]%
        {pysnooper}
\bibfield{author}{\bibinfo{person}{Ram Rachum}, \bibinfo{person}{Alex Hall},
  \bibinfo{person}{Iori Yanokura}, {et~al\mbox{.}}}
  \bibinfo{year}{2019}\natexlab{}.
\newblock \bibinfo{booktitle}{\emph{PySnooper: Never use print for debugging
  again}}.
\newblock
\urldef\tempurl%
\url{https://doi.org/10.5281/zenodo.10462459}
\showDOI{\tempurl}


\bibitem[\protect\citeauthoryear{Ramírez}{Ramírez}{2018}]%
        {fastapi}
\bibfield{author}{\bibinfo{person}{Sebastián Ramírez}.}
  \bibinfo{year}{2018}\natexlab{}.
\newblock \bibinfo{booktitle}{\emph{{FastAPI}}}.
\newblock
\urldef\tempurl%
\url{https://github.com/tiangolo/fastapi}
\showURL{%
\tempurl}


\bibitem[\protect\citeauthoryear{Roztocil and {contributors to
  Httpie}}{Roztocil and {contributors to Httpie}}{2012}]%
        {httpie}
\bibfield{author}{\bibinfo{person}{Jakub Roztocil} {and}
  \bibinfo{person}{{contributors to Httpie}}.} \bibinfo{year}{2012}\natexlab{}.
\newblock \bibinfo{title}{Httpie}.
\newblock
\newblock
\urldef\tempurl%
\url{https://github.com/jakubroztocil/httpie}
\showURL{%
\tempurl}


\bibitem[\protect\citeauthoryear{Saha, Khurshid, and Perry}{Saha
  et~al\mbox{.}}{2014}]%
        {bugLifetimes2}
\bibfield{author}{\bibinfo{person}{Ripon~K. Saha}, \bibinfo{person}{Sarfraz
  Khurshid}, {and} \bibinfo{person}{Dewayne~E. Perry}.}
  \bibinfo{year}{2014}\natexlab{}.
\newblock \showarticletitle{An empirical study of long lived bugs}. In
  \bibinfo{booktitle}{\emph{2014 Software Evolution Week - {IEEE} Conference on
  Software Maintenance, Reengineering, and Reverse Engineering, {CSMR-WCRE}
  2014, Antwerp, Belgium, February 3-6, 2014}},
  \bibfield{editor}{\bibinfo{person}{Serge Demeyer}, \bibinfo{person}{Dave~W.
  Binkley}, {and} \bibinfo{person}{Filippo Ricca}} (Eds.).
  \bibinfo{publisher}{{IEEE} Computer Society}, \bibinfo{pages}{144--153}.
\newblock
\urldef\tempurl%
\url{https://doi.org/10.1109/CSMR-WCRE.2014.6747164}
\showDOI{\tempurl}


\bibitem[\protect\citeauthoryear{{Sanic Community Organization}}{{Sanic
  Community Organization}}{2017}]%
        {sanic}
\bibfield{author}{\bibinfo{person}{{Sanic Community Organization}}.}
  \bibinfo{year}{2017}\natexlab{}.
\newblock \bibinfo{title}{Sanic}.
\newblock
\newblock
\urldef\tempurl%
\url{https://github.com/sanic-org/sanic}
\showURL{%
\tempurl}


\bibitem[\protect\citeauthoryear{{Scrapy Developers}}{{Scrapy
  Developers}}{2012}]%
        {scrapy}
\bibfield{author}{\bibinfo{person}{{Scrapy Developers}}.}
  \bibinfo{year}{2012}\natexlab{}.
\newblock \bibinfo{title}{Scrapy}.
\newblock
\newblock
\urldef\tempurl%
\url{https://github.com/scrapy/scrapy}
\showURL{%
\tempurl}


\bibitem[\protect\citeauthoryear{Sobreira, Durieux, Madeiral, Monperrus, and
  de~Almeida~Maia}{Sobreira et~al\mbox{.}}{2018}]%
        {d4j_bug_identify}
\bibfield{author}{\bibinfo{person}{Victor Sobreira}, \bibinfo{person}{Thomas
  Durieux}, \bibinfo{person}{Fernanda Madeiral}, \bibinfo{person}{Martin
  Monperrus}, {and} \bibinfo{person}{Marcelo de Almeida~Maia}.}
  \bibinfo{year}{2018}\natexlab{}.
\newblock \showarticletitle{Dissection of a bug dataset: Anatomy of 395 patches
  from Defects4J}. In \bibinfo{booktitle}{\emph{25th International Conference
  on Software Analysis, Evolution and Reengineering, {SANER} 2018, Campobasso,
  Italy, March 20-23, 2018}}, \bibfield{editor}{\bibinfo{person}{Rocco
  Oliveto}, \bibinfo{person}{Massimiliano~Di Penta}, {and}
  \bibinfo{person}{David~C. Shepherd}} (Eds.). \bibinfo{publisher}{{IEEE}
  Computer Society}, \bibinfo{pages}{130--140}.
\newblock
\urldef\tempurl%
\url{https://doi.org/10.1109/SANER.2018.8330203}
\showDOI{\tempurl}


\bibitem[\protect\citeauthoryear{{The pandas development team}}{{The pandas
  development team}}{2010}]%
        {pandas}
\bibfield{author}{\bibinfo{person}{{The pandas development team}}.}
  \bibinfo{year}{2010}\natexlab{}.
\newblock \bibinfo{booktitle}{\emph{{pandas-dev/pandas: Pandas}}}.
\newblock
\urldef\tempurl%
\url{https://doi.org/10.5281/zenodo.3509134}
\showDOI{\tempurl}


\bibitem[\protect\citeauthoryear{{Tornado Developers}}{{Tornado
  Developers}}{2013}]%
        {tornado}
\bibfield{author}{\bibinfo{person}{{Tornado Developers}}.}
  \bibinfo{year}{2013}\natexlab{}.
\newblock \bibinfo{title}{Tornado Web Server}.
\newblock
\newblock
\urldef\tempurl%
\url{https://github.com/tornadoweb/tornado}
\showURL{%
\tempurl}


\bibitem[\protect\citeauthoryear{Vokolos and Frankl}{Vokolos and
  Frankl}{1998}]%
        {space}
\bibfield{author}{\bibinfo{person}{Filippos~I. Vokolos} {and}
  \bibinfo{person}{Phyllis~G. Frankl}.} \bibinfo{year}{1998}\natexlab{}.
\newblock \showarticletitle{Empirical Evaluation of the Textual Differencing
  Regression Testing Technique}. In \bibinfo{booktitle}{\emph{1998
  International Conference on Software Maintenance, {ICSM} 1998, Bethesda,
  Maryland, USA, November 16-19, 1998}}. \bibinfo{publisher}{{IEEE} Computer
  Society}, \bibinfo{pages}{44--53}.
\newblock
\urldef\tempurl%
\url{https://doi.org/10.1109/ICSM.1998.738488}
\showDOI{\tempurl}


\bibitem[\protect\citeauthoryear{Widyasari, Sim, Lok, Qi, Phan, Tay, Tan, Wee,
  Tan, Yieh, Goh, Thung, Kang, Hoang, Lo, and Ouh}{Widyasari
  et~al\mbox{.}}{2020}]%
        {bugsinpy}
\bibfield{author}{\bibinfo{person}{Ratnadira Widyasari},
  \bibinfo{person}{Sheng~Qin Sim}, \bibinfo{person}{Camellia Lok},
  \bibinfo{person}{Haodi Qi}, \bibinfo{person}{Jack Phan},
  \bibinfo{person}{Qijin Tay}, \bibinfo{person}{Constance Tan},
  \bibinfo{person}{Fiona Wee}, \bibinfo{person}{Jodie~Ethelda Tan},
  \bibinfo{person}{Yuheng Yieh}, \bibinfo{person}{Brian Goh},
  \bibinfo{person}{Ferdian Thung}, \bibinfo{person}{Hong~Jin Kang},
  \bibinfo{person}{Thong Hoang}, \bibinfo{person}{David Lo}, {and}
  \bibinfo{person}{Eng~Lieh Ouh}.} \bibinfo{year}{2020}\natexlab{}.
\newblock \showarticletitle{BugsInPy: a database of existing bugs in Python
  programs to enable controlled testing and debugging studies}. In
  \bibinfo{booktitle}{\emph{{ESEC/FSE} '20: 28th {ACM} Joint European Software
  Engineering Conference and Symposium on the Foundations of Software
  Engineering, Virtual Event, USA, November 8-13, 2020}},
  \bibfield{editor}{\bibinfo{person}{Prem Devanbu}, \bibinfo{person}{Myra~B.
  Cohen}, {and} \bibinfo{person}{Thomas Zimmermann}} (Eds.).
  \bibinfo{publisher}{{ACM}}, \bibinfo{pages}{1556--1560}.
\newblock
\urldef\tempurl%
\url{https://doi.org/10.1145/3368089.3417943}
\showDOI{\tempurl}


\bibitem[\protect\citeauthoryear{Zheng, Wang, Fan, Chen, and Yang}{Zheng
  et~al\mbox{.}}{2018}]%
        {DBLP:journals/jss/ZhengWFCY18}
\bibfield{author}{\bibinfo{person}{Yan Zheng}, \bibinfo{person}{Zan Wang},
  \bibinfo{person}{Xiangyu Fan}, \bibinfo{person}{Xiang Chen}, {and}
  \bibinfo{person}{Zijiang Yang}.} \bibinfo{year}{2018}\natexlab{}.
\newblock \showarticletitle{Localizing multiple software faults based on
  evolution algorithm}.
\newblock \bibinfo{journal}{\emph{J. Syst. Softw.}}  \bibinfo{volume}{139}
  (\bibinfo{year}{2018}), \bibinfo{pages}{107--123}.
\newblock
\urldef\tempurl%
\url{https://doi.org/10.1016/j.jss.2018.02.001}
\showDOI{\tempurl}


\end{thebibliography}

\end{document}